\documentclass[aps, reprint, superscriptaddress, floatfix, noeprint]{revtex4-1}

\usepackage{hyperref}
\hypersetup{
    colorlinks=true,
    linkcolor=blue,
    citecolor=red,      
    urlcolor=blue,
    pdfpagemode=FullScreen,
    }
\usepackage{graphicx}
\usepackage{amsmath}
\usepackage{wasysym}
\usepackage{amsfonts}
\usepackage{color}
\usepackage{verbatim}
\usepackage{upgreek}
\usepackage{bm}
\usepackage{textcomp}
\usepackage[symbol*]{footmisc}
\usepackage{float}
\usepackage{siunitx}
\usepackage{soul}

\usepackage[utf8]{inputenc}
\usepackage[T1]{fontenc}
\usepackage{mathptmx}
\usepackage{etoolbox}

\DefineFNsymbolsTM{otherfnsymbols}{
  \textdagger    \dagger
  \textdaggerdbl \ddagger
  \textsection   \mathsection
  \textparagraph \mathparagraph
 }
\setfnsymbol{otherfnsymbols}

\newcommand{\fig}[1]{Fig.\,\ref{#1}}

\begin{document}

\title[]{Hyperspectral photoluminescence and reflectance microscopy of 2D materials} 

\author{David Tebbe} 
\email{david.tebbe@rwth-aachen.de}
\affiliation{2nd Institute of Physics and JARA-FIT, RWTH Aachen University, Aachen, Germany}
\author{Marc Schütte} 
\affiliation{2nd Institute of Physics and JARA-FIT, RWTH Aachen University, Aachen, Germany}
\author{Baisali Kundu} 
\affiliation{Materials Science Center, Indian Institute of Technology, Kharagpur, West Bengal, India}
\author{Bernd Beschoten} 
\affiliation{2nd Institute of Physics and JARA-FIT, RWTH Aachen University, Aachen, Germany}
\affiliation{JARA-FIT Institute for Quantum Information, Forschungszentrum J\"ulich GmbH and RWTH Aachen University, Aachen, Germany}
\author{Prasana K. Sahoo}
\affiliation{Materials Science Center, Indian Institute of Technology, Kharagpur, West Bengal, India}
\author{Lutz Waldecker} 
\email{waldecker@physik.rwth-aachen.de}
\affiliation{2nd Institute of Physics and JARA-FIT, RWTH Aachen University, Aachen, Germany}

\begin{abstract}

Optical micro-spectroscopy is an invaluable tool for studying and characterizing samples ranging from classical semiconductors to low-dimensional materials and heterostructures. 
To date, most implementations are based on point-scanning techniques, which are flexible and reliable, but slow. 
Here, we describe a setup for highly parallel acquisition of hyperspectral reflection and photoluminescence microscope images using a push-broom technique.
Spatial as well as spectral distortions are characterized and their digital corrections are presented. 
We demonstrate close-to diffraction-limited spatial imaging performance and a spectral resolution  limited by the spectrograph.
The capabilities of the setup are demonstrated by recording a hyperspectral photoluminescence map of a CVD-grown MoSe$_2$-WSe$_2$ lateral heterostructure, from which we extract the luminescence energies, intensities and peak widths across the interface. 

\end{abstract}

\maketitle

\section{Introduction}

Optical spectroscopy has proven to be an essential tool in developing the field of two-dimensional semiconductors.
Starting with the observation of an indirect-to-direct band gap transition \cite{Mak2010, Splendiani2010}, it has led, among many others, to the observation of the valley-dependent selection rules \cite{Cao2012, Ersfeld2020}, large many-body excitonic states \cite{Chen2018, Ye2018}, excitons in moir\'e potentials \cite{Jin2019} and magnetic proximity couplings \cite{Seyler2018}. 

Spectroscopy in the visible and near-infrared spectral region can be performed with sub-micrometer resolution if the light is focused down to (or near)  diffraction-limited spots. 
This is of particular importance for samples with locally varying properties, such as lateral heterojunctions \cite{Huang2014}, (gate-defined) nano-scale potentials for exciton confinement \cite{Palacios-Berraquero2017, Thureja2022}, or heterostructures of different materials or stacking orders \cite{Bru-Chevallier2006, Tebbe2023} as well as for material-specific inhomogeneities, such as strain, charge and screening-induced disorder \cite{Shin2016, Raja2019, Kolesnichenko2020} or localized defects \cite{He2015}. 

To visualize and characterize such features, hyperspectral images, i.e. data cubes that contain data points in two spatial and one spectral dimensions, can be acquired across the entire sample. 
Most often, this is implemented as point-scanning techniques in which light is focused to a tight spot and scanned across the sample, either by translating the sample with mechanical stages or by moving the focus using galvanic mirrors. 
While this technique is straight forward to implement, it can become very time-consuming if sample areas are large and need to be scanned at high spatial resolution or if external parameter spaces, such as gate voltages or magnetic fields, are to be varied. 

Several pathways exist to increase the speed of the acquisition of hyperspectral images \cite{Wang2017, Candeo2019}. 
A common approach is to directly image the sample through different (variable) spectral filters. 
Although this can be a particularly fast method for multi-spectral imaging and works for photoluminescence and absorption imaging, achieving a spectral resolution over a wide range of wavelengths would require a large set of narrow-band filters or a highly tunable filter, which does not induce aberrations. 
A second approach is to measure several points at a time by using a line of light on the sample and a two-dimensional detector array, such as a CCD or CMOS camera, connected to a spectrometer. 
This technique requires mechanically scanning the sample in one dimension (also called push-broom), but one naturally achieves a high spectral sampling. 
While it has been applied to image biological samples in reflection \cite{Huebschman2002, Ortega2019}, these implementations were not designed for conducting different optical experiments, such as photoluminescence spectroscopy, quantitative reflectance measurements or to measure polarization states.
The combination of several different experiments, however, is typically needed to obtain a clear picture of the optical transitions and processes in 2D materials \cite{Mak2010, Niehues2020}. 

Here, we present a setup for push-broom hyperspectral reflection and photoluminescence (PL) imaging, which is able to acquire hyperspectral images at diffraction-limited performance and a spectral resolution limited by the spectrograph, which can be extended to include polarization optics.
We discuss aberrations from the setup and digital image corrections, which allow the use of our setup for quantitative data analysis. 

\section{Setup}
\begin{figure}[ht!]
    \begin{center}
        \includegraphics[width=\linewidth]{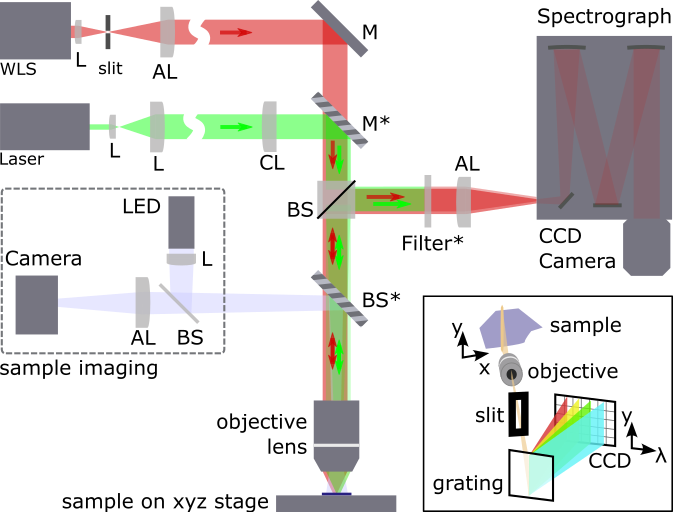}
        \caption{
        Design of the tabletop hyperspectral imaging setup for reflection (red beam path) or photoluminescence (green beam path) experiments.
        A linefocus is created by using a combination of two lenses and a slit in the white light beam path and by a cylindrical lens in the collimated laser beam path. 
        Inset: Sketch of the linefocus on the sample, which is imaged onto the entrance slit of the spectrograph. This creates a two-dimensional image on the CCD with one spatial dimension and one spectral dimension. 
        WLS: white light source, M: mirror, AL/CL/L: achromatic lens/ cylindrical lens / lens, BS: beam splitter.
        Optics marked with stars can be flipped out of the beam path to swap between the different measurement modes. 
        }
        \label{fig:setup}
    \end{center}
\end{figure}

The setup presented here is designed to be used for both, white light reflection, as well as for photoluminescence imaging in the visible and near-infrared range (400-800 nm). 
In both modes, a line-focus is created on the sample. 
The reflected or emitted light from this line-shaped area is projected onto the entrance slit of an imaging spectrometer.
In the spectrometer, this creates a two-dimensional image on the CCD camera, with one dimension corresponding to wavelength and the other corresponding to a spatial dimension.
After the acquisition of such an image, the sample is moved along an axis perpendicular to the line-focus by a motorized mechanical stage. 
An overview of the setup, which is placed on an optical table, is shown in \fig{fig:setup}.

The setup is equipped with several light sources, which can be changed using flip mirrors, including a 250 W tungsten-halogen lamp for white light experiments and a 40 mW, 532 nm continuous wave (cw) laser (Thorlabs DJ532-40).
The line-focus is created in two different ways for these light sources.
Since the white light source is spatially extended, it needs to be spatially filtered in order to achieve a tight focus.
In point-scanning setups, this is typically achieved with a pinhole in the common focus plane of two lenses. 
Here, the pinhole is replaced by a slit of dimension 20 \textmu m $\times$ 3 mm in between two achromatic doublet lenses.
The second lens collimates the white light, which is focused onto the sample using an objective lens. 
In our setup, several objectives are available, including a 100$\times$, NA=0.9 objective. 
The dimension of the slit and the collimating lens are chosen such that the image of the slit on the sample is just below the resolution limit of the lens.
This ensures that the width of the line is as small as it can be while maintaining an intensity high enough for fast image acquisition. 

For PL measurements, different laser sources can be employed; here, we use a 532 nm continuous wave (cw) laser (Thorlabs DJ532-40).
It is first expanded to exceed the dimension of the back aperture of the microscope objective. 
The laser beam is then passed through a cylindrical lens ($f_x = \infty $, $f_y = $ 40 cm), which is placed in front of the beam splitter, at a distance of approximately $f_y$ in front of the microscope objective.
As the objective only focuses the collimated dimension, a line-focus is generated on the sample plane.

The reflected or emitted light from the line-shaped area on the sample is collected and collimated by the objective.
It is then split from the incoming light path by a non-polarizing beam slitter and imaged onto the slit of the imaging spectrometer (Andor Shamrock 500i with gratings of 150, 300 and 600l/mm) by an achromatic lens ($f = $ 15 cm).
Optional filters or polarization optics can be introduced in front of the focusing lens. 
Inside the spectrometer, the line is split into its wavelengths by the grating, creating a two-dimensional image on the CCD camera (Andor Newton DU920P-BEX2-DD).
The vertical-direction of the CCD-images corresponds to a real-space y-dimension, and the horizontal-direction to a wavelength-space $\lambda$-dimension. 

Finally, a hyperspectral data cube is recorded by moving the sample by a motorized xyz stage perpendicular to the long focus direction while taking spectral images at each position (x and y motors: PI V-408, z stage: Newport M-MVN80 with TRA25PPD actuator). 
Since some level of chromatic aberration exists in the setup, the raw images need to be digitally corrected before the data can be analyzed. 

For alignment purposes, both the sample and the focus can be viewed simultaneously using an imaging setup.
This imaging module consists of a lighting arm with an LED Light Source (Thorlabs, MNWHL4) and an aspheric collimation lens, a beam splitter, and the imaging arm with a digital color camera (The imaging source, DFK 33UX265) and an achromatic tube lens ($f=$ 10 cm). 
The imaging setup is introduced into the beam path with a pellicle beam splitter, which keeps the offset of the light sources on the sample to a minimum.

\section{Image analysis}
\subsection{Chromatic aberrations}
Due to the use of refractive optics, some level of chromatic aberration is present in the setup.
This aberration manifests mainly as a wavelength dependence of the spatial dimension on the CCD camera of the imaging spectrometer.

\begin{figure*}[ht!]
    \begin{center}
        \includegraphics[width = 1.0\linewidth]{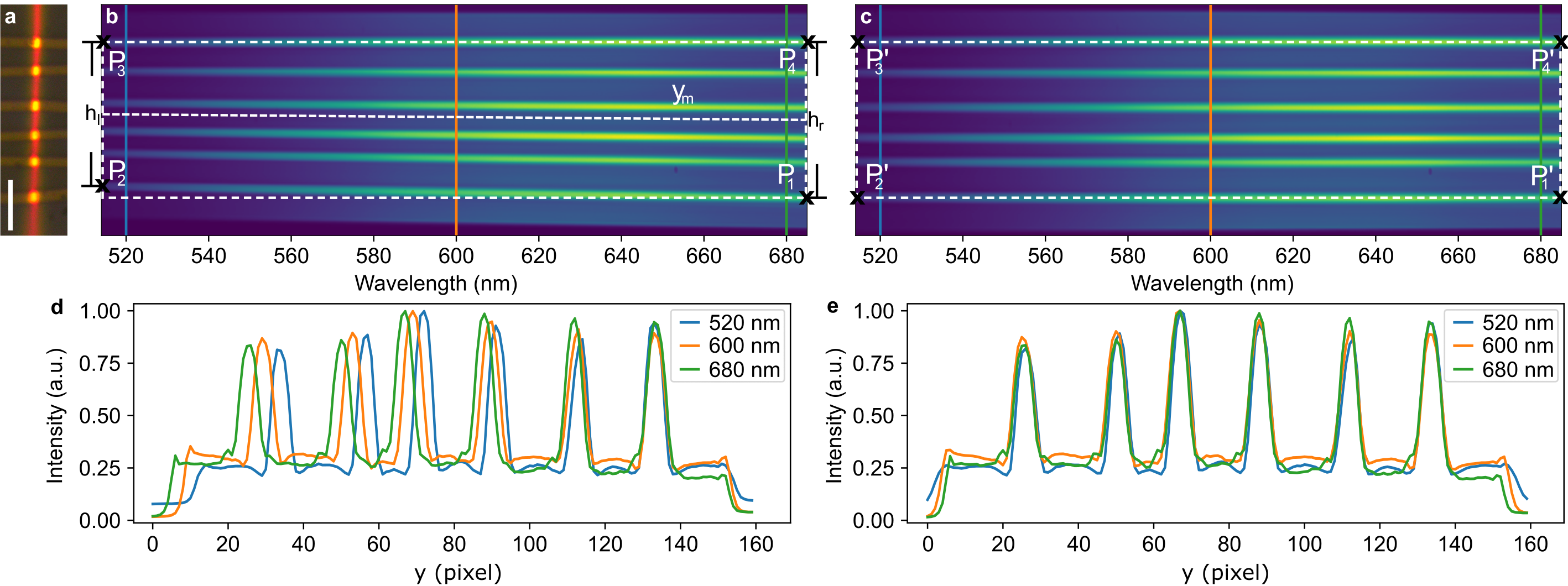}
        \caption{a) Optical microscope image of evaporated gold stripes on Si/SiO$_2$ with the white light focus vertically illuminating the structure. Scalebar: 5 \textmu m. 
        b) Raw image of the reflected white light of the structure taken by a CCD camera of the spectrograph. The x direction covers the wavelengths from 515 nm to 685 nm and the y direction corresponds to the direction along the focal line. A distortion in the y direction is visible due to chromatic aberrations of the setup.
        The white dashed horizontal lines help visualize the spectral distortion of the image.  
        c) Image after applying a digital remapping, which removes the aberration visible in b.
        d) Line cuts of the raw CCD-image at 520 nm, 600 nm and 680 nm, respectively.
        e) Line cuts of the remapped image at the same wavelengths, which show that the distortions have been largely removed. }
        \label{fig:imcorr}
    \end{center}
\end{figure*}

To visualize the distortions, we take white light reflectance images from a calibration sample of micro-patterned stripes of 5/50 nm thick Cr/Au with 0.8 \textmu m width and unequal spacing, fabricated on Si/SiO$_2$. 
\fig{fig:imcorr} a shows a microscope image of the white light line-focus on this sample.
The corresponding spectral image, taken with the spectrograph, is shown in \fig{fig:imcorr} b.
The vertical direction corresponds to the spatial dimension along the line-focus. 
Since gold has a higher reflectance than Si/SiO$_2$ across this wavelength range, it appears as bright lines in the spectral dimension. 

The lines in the spectral image are tilted, which is also seen in the different sizes of the vertical cuts at three different wavelengths, plotted in Figure \ref{fig:imcorr} d.
These distortions need to be corrected in order to retrieve the actual reflectance (or photoluminescence) spectra at each spatial point from horizontal cuts of the image. 

To quantify the distortions, we first find four coordinates $P_i = (x_i, y_i)$ in the image, which correspond to two spatial points on the sample at two different wavelengths.
As the wavelength axis is aberration corrected by the spectrograph, $x_1 = x_4$ and $x_2 = x_3$.
The y-coordinates can be found from the vertical line cuts, e.g. by the intensity maximum of the topmost and bottom-most gold feature as shown in \fig{fig:imcorr} d. 
We find that the distortions can be well approximated to be linear in wavelength, so that the four coordinates generally lie on a (nonsymmetric) trapezoid. 
In the corrected image, these four coordinates need to form a rectangle, i.e. $y_3 - y_2 = y_4 -y_1$ and $y_2 = y_1$.

For the image transformation, we work backwards, starting from a regular rectangular grid into which the raw data is supposed to be transformed.
This grid is squeezed and sheared in y-direction to match the shape of the trapezoid, creating a grid of unequal spacing, but of the same amount of points at each wavelength. 
The intensities of the measured image are then interpolated at the positions of this grid.
If displayed as an array of equally spaced pixels, this grid corresponds to the corrected image. 
Importantly, this transformation is different from an affine transform, as the x direction (wavelength) is not modified in this transformation.

The transformation of the rectangular grid is performed in the following way: first, the grid points are squeezed in y-direction by a factor $s(x)$.
Using the side lengths $h_l = |y_3 - y_2|$ and $h_r = |y_1 - y_4|$, the squeezing factor is given by:
\begin{equation}
    s(x) = \frac{h_l}{h_r} + \left( 1- \frac{h_l}{h_r} \right)\frac{x}{x_r-x_l}
\end{equation}
Secondly, it is sheared such that the center line matches the middle line of the distorted image $y_m(x)$. 
The position of the middle line $y_m(x)$ is a linear function, given by the middle heights on each side $y_m(x_r) = \frac{y_1 + y_4}{2} $ and $y_m(x_l) = \frac{y_2 + y_3}{2} $:
\begin{equation}
    y_m(x) = y_m(x_l) +\left( y_m(x_r) - y_m(x_l) \right) \frac{x}{x_r-x_l}
\end{equation}

The  coordinates of each pixel on the distorted grid, $y(x)$, are then given in terms of the positions on the rectangular grid, $y'(x)$, by:
\begin{equation}
    y(x) = s(x) \left( y'(x) - y_m(x_r) \right) + y_m(x)
\end{equation}
The interpolation of the image onto these coordinates is implemented using the \textit{remap} function of the python package \textit{OpenCV}.
The resulting image and line cuts are shown in \fig{fig:imcorr} c and e. 
A horizontal line cut now corresponds to the reflected spectrum at one particular spatial point. 
The transformation is generally applied to all images before further analysis.
A new calibration must be performed after each major realignment of the setup. 

\subsection{Intensity profiles}

Images of the line-foci as well as their spatial profiles along the x and y directions are shown in \fig{fig:foci}.
Along the long dimension (x-direction), the white light intensity is relatively flat over several tens of micrometers, but contains some features. 
The laser has an almost Gaussian shape with a full width at half maximum (FWHM) of 16 \textmu m. 
These profiles directly result from the different ways the lines are achieved: as the intensity along the slit is homogeneous, so is the resulting focus, whereas the profile of the laser originates from the Gaussian shape of the collimated laser beam. 
Note that using a slit in the collimated laser beam instead of a cylindrical lens does not yield a more homogeneous intensity profile along the long direction, but rather leads to fringes due to diffraction \cite{Luo2015Oct}.

For a quantitative analysis of the data, the varying intensity of both white light and laser needs to be taken into account. 
In white light experiments, one often determines the reflection contrast, i.e. the reflected intensity on the sample divided by the intensity next to the sample.
This eliminates the parameters of the setup, such as the emission spectrum of the lamp and sensitivity curve of the detector. 
It is possible, as the intensity of the light is typically low enough to neglect nonlinear effects. 
Using the push-broom technique introduced here, the reflection contrast can either be calculated taking a reference on a single point, which introduces a small error due to the remaining fluctuations of the white light. 
It is therefore favorable to use an entire reference line, which further eliminates these variations (see Supplementary Information for details on the implementation).

For photoluminescence experiments, the excitation power varies to a larger degree due to the profile of the laser.
By taking a reference measurement of the laser intensity on the CCD of the spectrograph (or the PL intensity of a homogeneous test sample), it can nevertheless be accounted for.
In this case, each row of the acquired data images is divided by the (normalized) intensity of the reference measurement. 
Note that this approach requires that the PL intensity be linearly dependent on the excitation power.
Although this holds well, for example for neutral excitons in transition metal dichalcogenides (TMDs), it might not for the emission of other states, such as from localized defects \cite{Tongay2013} or biexcitons \cite{Ye2018, Chen2018}.
To study states with a nonlinear excitation dependence with the push-broom technique, the intensity in the y direction would need to be homogenized, e.g. by the use of a top-hat beam shaper or by selecting only a small fraction of the line. 
In order to stay within reasonable variations of excitation density, we typically truncate the PL images at approximately 50\% of the maximum laser intensity and apply the intensity correction to all images.

\section{Resolution of the setup}

\begin{figure}[ht!]
    \begin{center}
        \includegraphics[width=\linewidth]{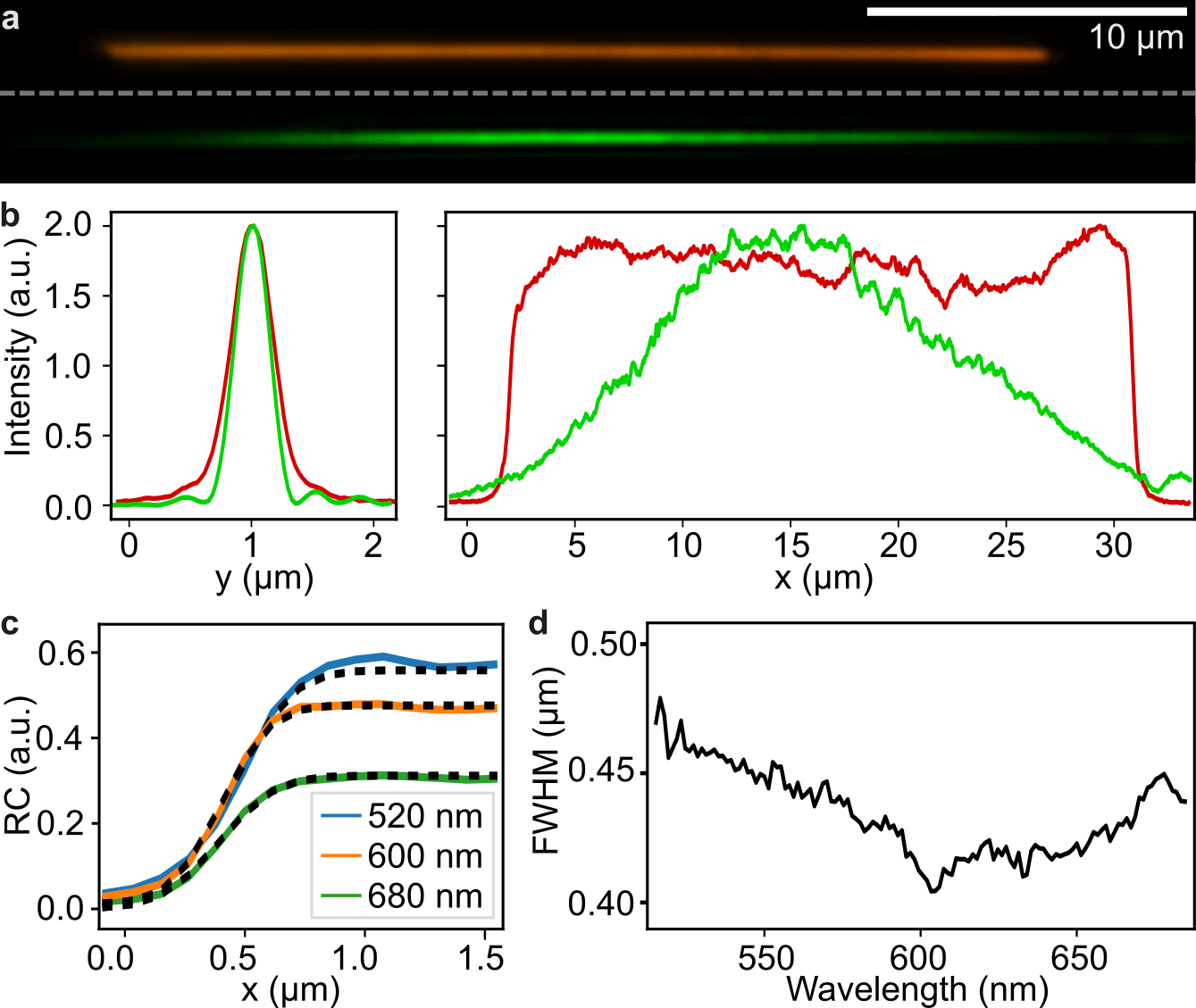} 
        \caption{
        a) Optical images of the focused white-light source (upper) and the 532 nm laser focus (lower).
        b) Vertical (left) and horizontal (right) intensity profiles of the foci of the white light (red) and laser (green). The vertical FWHM sizes extracted are 0.4 \textmu m (white light) and 0.31 \textmu m (laser). 
        In the horizontal direction, the white light has a flat profile over a length of approximately 30 \textmu m while the laser has an approximately Gaussian profile with a FWHM of  16 \textmu m.
        c) Reflection contrast measurements with the white light source, using the edge of a trilayer graphene as a knife edge. An error function is fitted (black dashed lines) to extract the FWHM and thus the spatial resolution along the line-focus.
        d) FWHM extracted from one knife-edge image, evaluated at wavelengths between 515 nm and 685 nm
        }
        \label{fig:foci}
    \end{center}
\end{figure}

We now discuss the performance of the setup. 
The spatial resolution perpendicular to the line-focus is trivially given by the width of the line-focus (see \fig{fig:foci} b). 
In this dimension (x-direction), we measure a FWHM of 0.40 \textmu m for the white light source (spectrally filtered at $620\pm20$ nm) and 0.31 \textmu m for the 532 nm laser. 
The slightly smaller width of the laser is likely related to the different beam profiles and the shorter wavelength of the laser beam.

To verify that the spatial resolution in the direction parallel to the line is comparable to the FWHM in perpendicular direction, we performed knife-edge measurements, in which the white light focus is placed over the edge of a trilayer graphene flake. 
The measured spectral image is analyzed by taking vertical cuts at each wavelength.
Three examples of these cuts are shown in \fig{fig:foci} d. 
They are analyzed by fitting an error-function to the data; the resulting FWHM corresponds to the spatial resolution of the setup and is shown in \fig{fig:foci} d). 
At 620 nm, the spatial resolution is indeed similar to the measured FWHM in the perpendicular direction. 
While the spatial resolution varies with wavelength, this is not a feature of the line-focus.
It results from the chromatic aberration of the lenses used in the setup and would also be present in point-scanning techniques. 
The wavelength of the best (i.e. diffraction-limited) spatial resolution can be adjusted by changing the position of the last focusing lens before the spectrometer. 

The spectral resolution of our setup is given by the resolution of the spectrograph. 
Due to the use of curved mirrors with an aberration corrected shape, it does not significantly vary with the y position (see the datasheet of the manufacturer). 
Using the grating with the highest groove density, the spectral resolution is on the order of 0.15 nm.

\section{PL imaging of lateral heterostructures}

\begin{figure*}[ht!]
    \begin{center}
        \includegraphics[width = 1\linewidth]{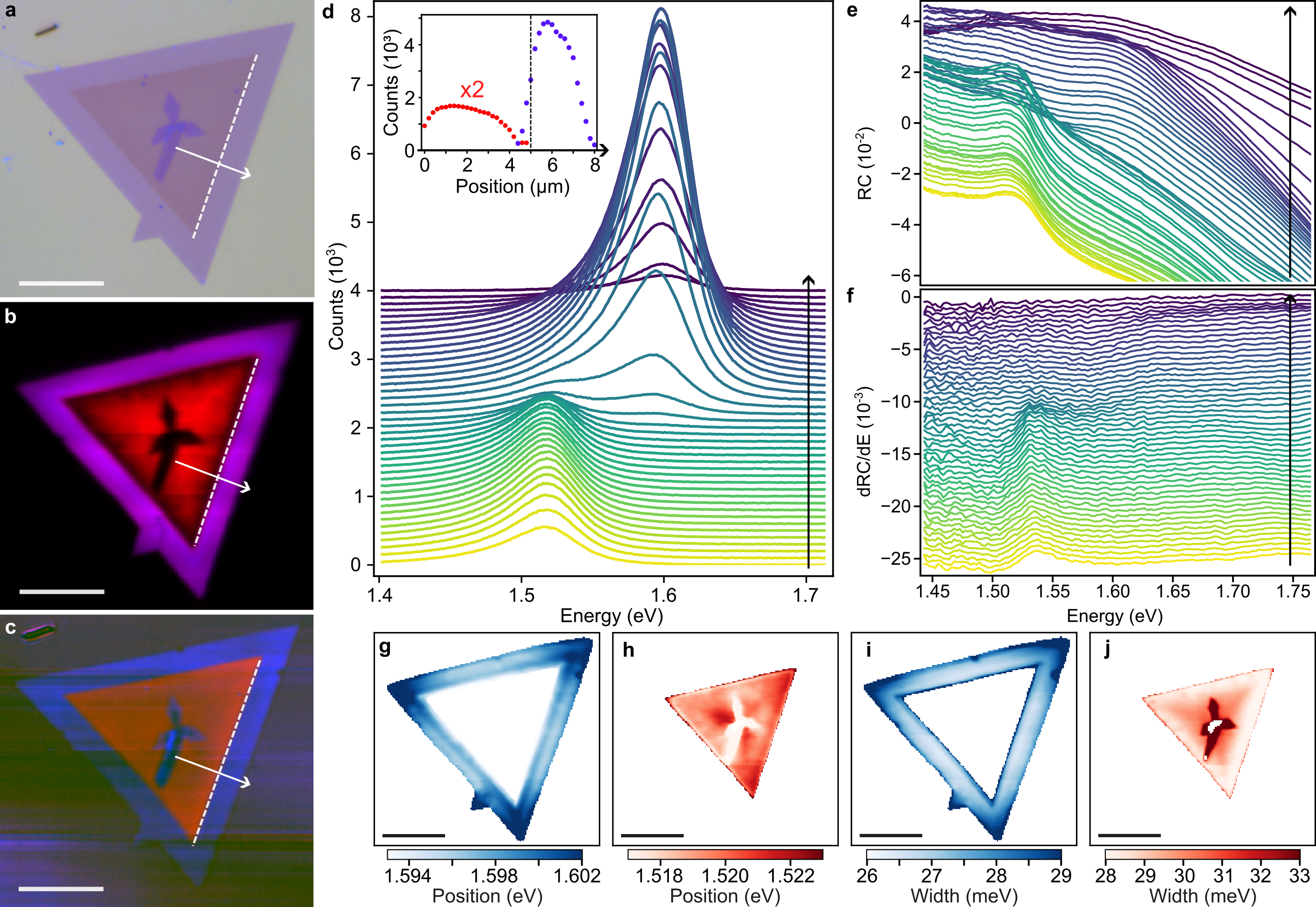}
        \caption{ a) Microscope image of a CVD-grown MoSe$_2$-WSe$_2$ lateral heterostructure. b) False-color photoluminescence image and c) false-color reflection contrast image of the same sample taken with the setup described in the main text. d) PL-Spectra extracted along the line perpendicular to the MoSe$_2$-WSe$_2$ interface, as indicated by the arrow in b. The inset shows the intensity across the interface of the MoSe$_2$ peak (red) and the WSe$_2$ peak (blue). e) RC-Spectra and f) its first derivative with respect to energy, extracted along the line perpendicular to the MoSe$_2$-WSe$_2$ interface, as indicated by the arrow in c. g,h) Peak position of WSe$_2$ (blue) and MoSe$_2$ (red), obtained from a fit of two Gaussians to the PL spectra. i,j) Corresponding peak widths within both areas. All scale bars are 10 \textmu m.}
        \label{fig:measurement}
    \end{center}
\end{figure*}

We demonstrate the capability of the setup by acquiring a hyperspectral photoluminescence and reflection contrast image of a TMD monolayer sample at room temperature; further reflection contrast measurements taken with the setup can be found in the study of Ref. \cite{Tebbe2023}.
The sample is a MoSe$_2$-WSe$_2$ in-plane heterostructure grown by chemical vapor deposition using the one-pot chemical vapor deposition growth method \cite{Sahoo2018}.
A micrograph of the sample is shown in \fig{fig:measurement} a.

Spectral images were taken with a laser power of 130~\textmu W ($\lambda=532 $nm), which translates to an irradiance of approximately 3 \textmu W/\textmu m$^2$ (note that the data were taken with a different cylindrical lens, which led to a laser profile of 7 \textmu m FWHM). 
To map the entire flake, 5 rows of 175 spectral images were taken, with step sizes of 7 \textmu m in the y direction and 0.2 \textmu m in the x direction.
Note that the step size in the x-direction is chosen to match the real-space pitch in the y-direction on the CCD, such that the images appear without distortions. 
Each image had an exposure time of 4 s and the total time to acquire all data, corresponding to 49000 individual spectra, was 75 min.
The 15 minutes not connected to data acquisition are linked to the communication with the camera and the moving of the mechanical stages.

The raw images are first corrected for chromatic aberration and excitation power as described in the previous sections. 
Afterwards, the five rows are stitched together and averaged where they overlap (see SI1 for details).
A background spectrum is subtracted by averaging an area without sample (see SI2 for details of background and reference spectra). 
A false-color image of the hyperspectral data is shown in \fig{fig:measurement} b, where the color channels are the integrated counts (normalized to their respective maximum) in the spectral range between 1.46 to 1.54 eV (red) and 1.56 to 1.64 eV (blue).
All major features seen in the micrograph can be identified (e.g. the thicker area in the middle), and the PL image appears without visible distortions.
Note that the WSe$_2$ areas appear purple, as the tail of the luminescence of WSe$_2$ reaches into the lower-energy integration window. 

In \fig{fig:measurement} c, a subset of the spectra of the data set are shown, extracted along a line perpendicular to the MoSe$_2$-WSe$_2$ interface (see the white arrow in \fig{fig:measurement} b).
In both, MoSe$_2$ and WSe$_2$ regions, the PL emission shows a single peak, which is slightly asymmetric in the WSe$_2$ region.
The intensity of MoSe$_2$ is significantly lower than that of WSe$_2$, which has been attributed to the different ordering of bright and dark states in the two materials \cite{Wang2015spin-orbit, Zhang2015}.

For a more quantitative analysis of the data, we fit the spectra with two Gaussian peaks.
The resulting peak-heights across the interface are shown in the inset of \fig{fig:measurement} c. 
The dashed line indicates the interface between the two materials as seen in the optical microscope image.
The most striking feature is that, while the intensity of WSe$_2$ drops right at the interface, the intensity of MoSe$_2$ is strongly reduced within a distance of approximately one micrometer from the interface.
The drop in intensity towards the interface could be caused by a higher defect density at the interface or by diffusion and charge carrier transfer to the WSe$_2$ side \cite{Berweger2020, Jia2020, Ambardar2022, Shimasaki2022, Beret2022}. 
At the used irradiance of 3 \textmu W/\textmu m$^2$, a reduction of PL due to exciton-exciton annihilation is not expected \cite{Lee2018Jul}.

The peak positions and widths of the entire sample are also determined by the fit and are color-coded in \fig{fig:measurement} d and e. 
Both peak energies are redshifted compared to literature values, indicating tensile strain across the entire sample \cite{Island2016, Aslan2018}, which is common for CVD grown materials on silica \cite{Ahn2017}.
Both the peak position and the width show variations of few meV across the respective materials. 
The most obvious feature in the WSe$_2$ is the shift towards higher peak energy and higher peak width at the corners of the triangle, which hints towards reduced tensile strain at the corners \cite{Aslan2018, Niehues2018}.
In the MoSe$_2$ area, a systematic change of the peak width can be observed, with the peaks getting narrower toward the outside, which points to a greater strain towards the interface \cite{Niehues2018}.

\section{Conclusion and outlook}

In conclusion, we implemented and characterized a hyperspectral push-broom micro-spectroscopy setup for highly parallel acquisition of reflection and photoluminescence spectra. 
To demonstrate the capabilities, we measured the PL spectra of a two-dimensional lateral heterostructure sample. 

The implementation of the parallel measurement scheme presented here can easily be implemented in existing setups and can be combined with various measurement techniques.
By inserting polarization optics, the setup can be used to map e.g. the polarization-resolved luminescence connected to the valley selection rules of TMDs \cite{Zeng2012,Mak2012}, or the magnetic circular dichroism in 2D heterostructures including magnetic materials \cite{Zhong2017_heterostructure, Seyler2018}. 
The concept can also be applied for Raman imaging, if a laser of sufficiently narrow line width in combination with appropriate filters is used and to other measurement techniques relevant to the semiconductor industry, such as optical critical dimension methods \cite{Hoobler2003}. 
Lastly, the setup can be modified to use reflective optics for a more broadband operation, potentially ranging from the infrared to the UV spectral region \cite{Ma2020}. 

The measurement scheme has broad applicability in studying quantitative optical properties of two-dimensional materials and beyond. It is particularly useful for imaging experiments of large sample areas or in situations in which multi-dimensional parameters scans are to be combined with high-resolution imaging.

\section*{Supplementary Material}
See the supplementary material for details on how multiple passes are stitched together and how background and reference spectra are extracted from the hyperspectral images, i.e. datacubes.

\section*{Data availability statement}
The data that support the findings of this study are openly available in zenodo, at \href{https://doi.org/10.5281/zenodo.7924667}{doi.org/10.5281/zenodo.7924667}.

\section*{Acknowledgments}
This project has received funding from the European Union's Horizon 2020 research and innovation program under grant agreement No. 881603 (Graphene Flagship). 
PKS acknowledges the Department of Science and Technology (DST), India (Project Code: DST/TDT/AMT/2021/003 (G)\&(C)).

\section*{References}

\end{document}


\title{Supplementary Information: Hyperspectral photoluminescence and reflectance microscopy of 2D materials} 

\author{David Tebbe} 
\email{david.tebbe@rwth-aachen.de}
\affiliation{2nd Institute of Physics and JARA-FIT, RWTH Aachen University, Aachen, Germany}
\author{Marc Schütte} 
\affiliation{2nd Institute of Physics, RWTH Aachen University, Aachen, Germany}
\author{Baisali Kundu} 
\affiliation{Materials Science Center, Indian Institute of Technology, Kharagpur, West Bengal, India}
\author{Bernd Beschoten} 
\affiliation{2nd Institute of Physics and JARA-FIT, RWTH Aachen University, Aachen, Germany}
\affiliation{JARA-FIT Institute for Quantum Information, Forschungszentrum J\"ulich GmbH and RWTH Aachen University, Aachen, Germany}
\author{Prasana Sahoo}
\affiliation{Materials Science Center, Indian Institute of Technology, Kharagpur, West Bengal, India}
\author{Lutz Waldecker} 
\email{waldecker@physik.rwth-aachen.de}
\affiliation{2nd Institute of Physics and JARA-FIT, RWTH Aachen University, Aachen, Germany}

\maketitle

\section{Supplementary Note 1: Stitching of measurements}

\begin{figure*}[bth]
    \begin{center}
        \includegraphics[width=1.0\textwidth]{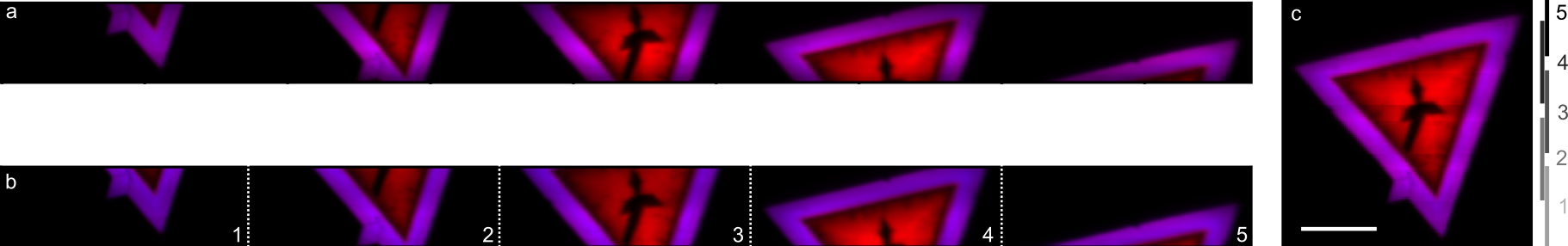}
        \caption{Image correction and stitching of multi-pass experiments. a) Raw image of the measurement shown in the main text. Five consecutive passes over the sample were taken with an offset of 7 \textmu m. 
        b) The same data as in a, after the intensity correction has been applied. 
        c) Image as shown in the main text after stitching the five passes together. }
        \label{fig:RCs}
    \end{center}
\end{figure*}

In both reflection and photoluminescence experiments, data is typically taken in several passes, i.e. at several different y-positions of the sample (see SI Figure \ref{fig:RCs} a). 
After applying image corrections, e.g. the intensity correction (SI Figure \ref{fig:RCs} b), the datacube is broken into smaller datacubes of the individual passes, which can then be stitched together. 
This is done by first determining the pitch of pixels in real-space. 
Multiplying the pitch with the step-size in y-direction known from the experiment, the offset in pixels between passes is calculated. 
The datacubes are then overlayed with this offset and overlapping pixels are averaged.

\section{Supplementary Note 2: Reference and background spectra}

In this paragraph, we describe how reference spectra are obtained, from which the reflection contrast is calculated in reflection measurements, and how background spectra are obtained and subtracted in photoluminescence (PL) experiments. 

\begin{figure*}[!tbh]
    \begin{center}
        \includegraphics[width=1.0\textwidth]{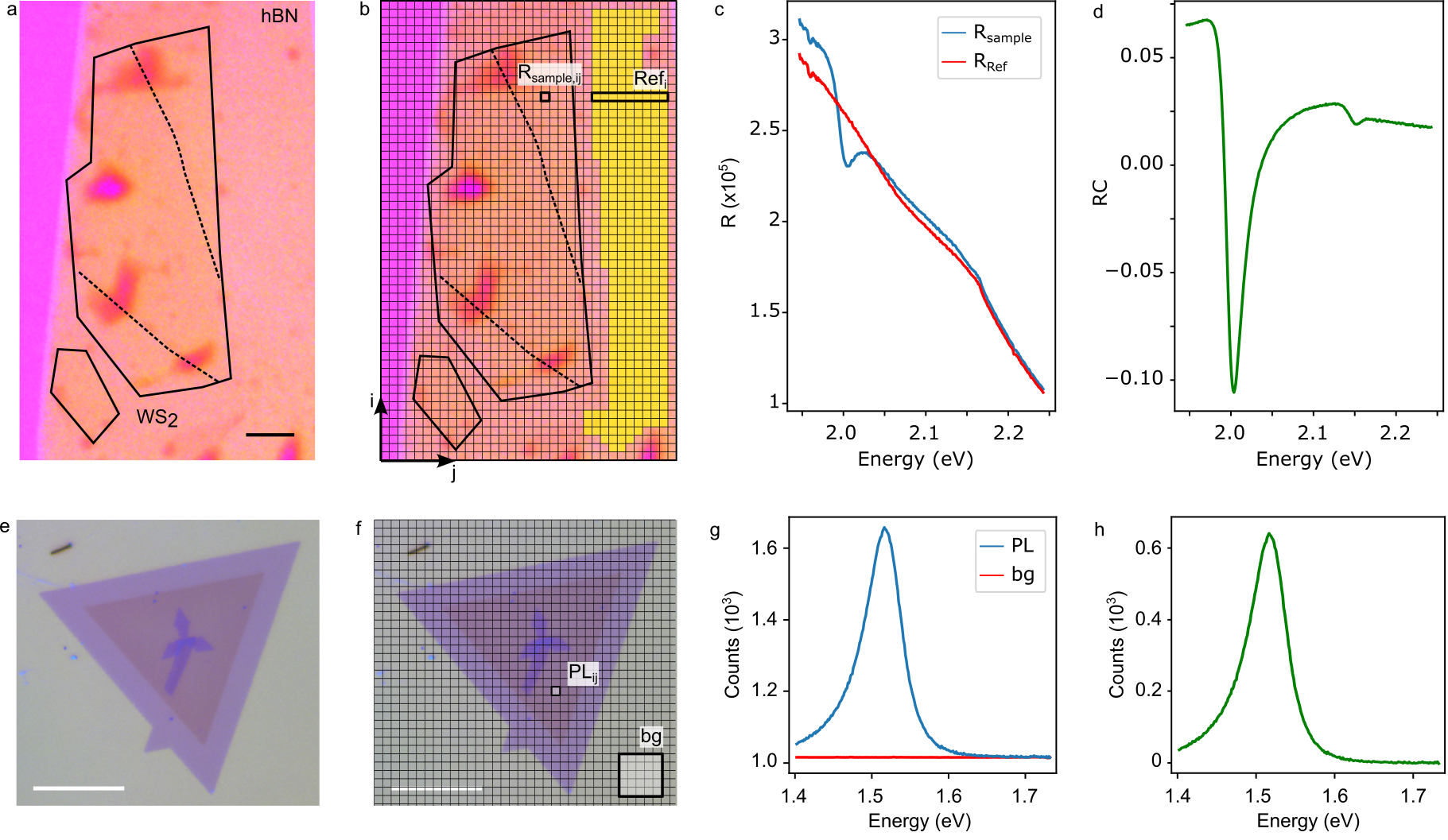}
        \caption{Extraction of Reference and Background spectra. a) Image of a WS2 encapsulated in hBN and additional graphene layers (for a detailed description of the sample see \cite{Tebbe2023}). b) Sketch of the extraction of reference spectra from a hyperspectral measurement. Reference spectra are calculated for each row. c) Raw spectrum on the sample $R_{sample}$ and the reference spectrum of the respective row. d) Reflection contrast calculated from the two spectra in c. 
        e) Image of the lateral TMD heterostructure described in the main text. f) Sketch of the extraction of background spectra, which are obtained by averaging an area next to the sample. g) PL spectrum on the sample (position as shown in f) and background spectrum.
        h) Background-corrected PL spectrum.
        }
        \label{fig:bg-ref}
    \end{center}
\end{figure*}

As the intensity along the whitelight linefocus can vary (see main text), it is favorable to use separate reference spectra $R_{\textrm{ref}, i}(\lambda)$ for each horizontal line $i$ of the image.
In practice, these spectra are obtained by first determining the area $A$ in which the reference is supposed to be taken, e.g. within a python script. 
The reference area is chosen to be of the same stacking order as the sample, but without the investigated material; an exemplary sketch with a possible reference area is shown in \ref{fig:bg-ref} b (the reference area is marked yellow). 
The individual reference spectra for each line are then calculated by averaging the spectra of each row within the area, $R_{\textrm{ref},i}(\lambda) = \sum_{j \in A} R_{ij}(\lambda)$.
Using this reference vector, the reflection contrast at each pixel of the hyperspectral image, $RC_{ij}(\lambda)$ is calculated as 
\begin{equation}
    RC_{ij}(\lambda) = R_{ij}(\lambda)/R_{\textrm{ref},i}(\lambda),
\end{equation}
where the division is done pixel-wise. 
A raw spectrum and its reference line are shown in SI Fig. \ref{fig:bg-ref} c and the corresponding reflection contrast is shown in \ref{fig:bg-ref} d.

In photoluminescence experiments, only a background needs to be subtracted. 
The background can be extracted from a measurement from an area $B$ without sample, see SI Figure \ref{fig:bg-ref}.
The background is simply calculated as $bg(\lambda) = \sum_{i,j \in B} PL_{ij}(\lambda)$.
An examplary raw PL spectrum and the averaged background are shown in SI Figure \ref{fig:bg-ref} g. 
The corrected spectrum is shown in SI Figure \ref{fig:bg-ref} h.

%